\documentclass[]{llncs}
\usepackage[T1]{fontenc}
\usepackage{graphicx}
\usepackage{hyperref}
\usepackage{float}
\usepackage{color}

\usepackage{subcaption}
\usepackage{enumitem}
\usepackage{cite}
\usepackage{amsmath}

\title{A Comparative Study of Text Retrieval Models on DaReCzech}
\def\nummodels{7}

\author{Jakub Stetina$^\diamondsuit$  \and
        Martin Fajcik$^\diamondsuit$  \and
        Michal Štefánik$^\clubsuit$ \and
        Michal Hradis$^\diamondsuit$\newline
        \vspace{10pt}\{xsteti05,ifajcik,ihradis\}@fit.vutbr.cz, stefanik.m@mail.muni.cz}

\authorrunning{}

\institute{$^\diamondsuit$Faculty of Information Technology, Brno University of Technology, Czech Republic \\
\vspace{5pt}$^\clubsuit$Faculty of Informatics, Masaryk University, Czech Republic}

\begin{document}
\maketitle              % typeset the header of the contrRibution
\begin{abstract}
This article presents a comprehensive evaluation of \nummodels{} off-the-shelf document retrieval models: Splade, Plaid, Plaid-X, SimCSE, Contriever, OpenAI ADA and Gemma2 chosen to determine their performance on the Czech retrieval dataset DaReCzech. The primary objective of our experiments is to estimate the quality of modern retrieval approaches in the Czech language. Our analyses include retrieval quality, speed, and memory footprint. Secondly, we analyze whether it is better to use the model directly in Czech text, or to use machine translation into English, followed by retrieval in English.
Our experiments identify the most effective option for Czech information retrieval. The findings revealed notable performance differences among the models, with Gemma22 achieving the highest precision and recall, while Contriever performing poorly. Conclusively, SPLADE and PLAID models offered a balance of efficiency and performance.
% pridat nieco o zaveroch, ktore sa nam podarilo urobit.
\end{abstract}

\section{Introduction}

Information retrieval (IR) is used in areas such as search engines and question-answering systems. Lately, we've seen advancements in IR models \cite[\textit{inter alia}]{hendrycks2021measuringmassivemultitasklanguage, yüksel2024turkishmmlumeasuringmassivemultitask, koto2024arabicmmluassessingmassivemultitask}, but picking the right one for a non-English document collection can be challenging. We address this gap for Czech language by doing a comprehensive comparison in our study. In particular, we utilize DareCzech, a Czech retrieval and ranking dataset \cite{kocián2021siamesebertbasedmodelweb}, for testing IR models to evaluate different IR models on Czech documents and queries. Our contributions are: (1) we analyze the index sizes to understand the storage requirements of various models, (2) we analyze the retrieval speed of such methods, to estimate how these models scale to large corpora in their default implementation, (3) we conduct ranking performance testing using multiple metrics on off-the-shelf models, and (4) we compare different model types, including those tested directly on the Czech dataset as well as on an English translation of the Czech dataset, keeping in mind the respective model's training data language, to provide insights into different approaches for indexing and retrieving czech. To the best of our knowledge, this is the first comparison study of existing state-of-the-art retrieval methods in the Czech language.

\section{Related Work}
Several well-known benchmarks have been used for evaluating information retrieval (IR) and text embedding models. \textbf{MS MARCO} \cite{bajaj2018msmarcohumangenerated} is widely used for passage and document retrieval, offering real-world web queries and answers. \textbf{MIRACL} \cite{zhang2022makingmiraclmultilingualinformation} is a multilingual benchmark designed for retrieval across different languages. \textbf{MTEB} \cite{muennighoff2023mtebmassivetextembedding} provides a comprehensive evaluation across diverse tasks, including clustering, classification, and re-ranking.

Beyond English-language benchmarks, several datasets focus on IR evaluation within specific linguistic contexts. For Czech, the \textbf{CWRCzech} dataset \cite{Von_sek_2024} includes 100M query-document pairs based on Czech click data from \emph{Seznam.cz} search logs. German-language IR  is explored through \textbf{DPR German} \cite{möller2021germanquadgermandprimprovingnonenglish} and \textbf{German LEGAL IR} \cite{wrzalik-krechel-2021-gerdalir}, which assess retrieval in general and legal domains. The \textbf{SKQuad} \cite{retrieval_skquad} dataset provides an IR benchmark specifically for the Slovak language and the \textbf{Scandinavian Embedding Benchmark (SEB)} \cite{enevoldsen2024scandinavianembeddingbenchmarkscomprehensive} provides a comprehensive evaluation framework for text embeddings in Scandinavian languages. Within MTEB, Polish \cite{poświata2024plmtebpolishmassivetext} and Chinese \cite{xiao2024cpackpackedresourcesgeneral} datasets extend the evaluation to language-specific IR tasks.

The dataset utilized in our study is \textbf{DaReCzech} (see Subsection \ref{sec:dareczech_dataset}), introduced in \cite{kocián2021siamesebertbasedmodelweb}, which is specifically tailored for the Czech language and consists of manually annotated query-document pairs. The relevance annotations in DaReCzech are not binary, allowing for a more nuanced evaluation of relevance ranking models and enabling the use of various evaluation metrics. 

\section{Model Descriptions} 
\begin{description}[style=unboxed,leftmargin=0em,listparindent=\parindent]
    \setlength\parskip{1em}
\item[BM25\cite{bm25}.]
BM25 is a traditional lexical approach which has been widely used and had been the standard before the rise of neural models. It ranks documents based on a query's term frequency, inverse document frequency, and document length, meaning the importance of each term in the query and document is considered along with the document's length normalization, to produce relevance scores for each document.

In our study, we employed a BM25 baseline to assess the effectiveness of the other models. This model stands out as the only model with lemmatization applied to the query and document content, a distinction arising from the nature of BM25, which is not a neural model and relies on the precise lexical form of terms within the corpus.

For the Czech language, which features word inflection, lemmatization is essential for precise term matching and relevance ranking. Therefore, the lemmatized version of the corpus for BM25 is required.

\item[splade-cocondenser-ensembledistil (SPLADE) \cite{formal2021spladesparselexicalexpansion}.]
    (Sparse Lexical and Expansion Model) leverages sparse vocabulary-sized representations to leverage the advantages of BOW (bag-of-words). Splade operates by first applying a linear transformation to the BERT \cite{devlin2019bertpretrainingdeepbidirectional} output embeddings, then performing a dot product with the token embeddings from the whole vocabulary, resulting in a matrix of scores where each input token has a score for every token in the vocabulary\footnote{It utilizes the already pretrained masked language modeling head.}. While its predecessor, SparTerm \cite{bai2020spartermlearningtermbasedsparse}, used a learned binary mask to select relevant scores from this matrix, Splade induces sparsity through a combination with a FLOPS regularizer and a logarithmic function during the representation computation. The final representation is obtained by summing the weights along the sequence tokens, producing a sparse embedding with a dimensionality equal to the vocabulary size. The second version of Splade improves this pooling mechanism to instead use the max for each token from the vocabulary. The model used in this comparison is the highest performing distilled version of splade as described in \cite{formal2022distillationhardnegativesampling}. 

\item[ColbertV2.0 (PLAID) \cite{santhanam2022plaidefficientenginelate}.]
The PLAID model represents a multi-vector approach. It extends the late interaction mechanism used in \textbf{ColBERT} \cite{DBLP:journals/corr/abs-2004-12832} to enhance efficiency in information retrieval. In the original version of ColBERT, the comparison of query-document embeddings was performed by matching every token in the document embedding with every token in the query embedding, calculating scores using a maximum similarity function, where the highest similarity value for each query token across all document tokens is retained. ColBERTv2 \cite{santhanam2022colbertv2effectiveefficientretrieval} improved upon this approach by clustering the document embeddings into centroid clusters, thereby enabling a more efficient retrieval process. At search time, a fixed number of candidate clusters are selected, and their embeddings are decompressed to compute the final similarity scores.

In addition, \textbf{PLAID}, further enhances efficiency and performance by introducing a multi-stage candidate generation process. This approach includes steps for pruning and centroid-based interactions, progressively narrowing down the set of candidate passages. The final, smaller set of potential passages is then scored, resulting in a more streamlined and scalable retrieval pipeline. The model used in this study was trained on the English MS MARCO. 

\item[Plaidx-xlmr-large-mlir-neuclir (PLAID-X) \cite{ecir2022colbert-x,ecir2024translate-distill}.]
Multilingual version of PLAID, PLAID-X builds upon the ColBERT architecture and employs a multilingual encoder XLM-RoBERTa (XLM-R) for multilingual and cross-lingual encodings. The model used in this study was trained using the translate-train approach on Chinese, Persian, and Russian data, relying on the XLM-R encoder for cross-language mappings.

\item[Text-embedding-ada-002 (OpenAI ADA) \cite{openai2023ada}.]
A closed-source model, that uses cosine similarity to compare two embeddings to calculate the resulting score.

\item[Contriever-msmarco (Contriever) \cite{gao2022simcsesimplecontrastivelearning}.]
Contriever is a dense retrieval model that leverages self-supervised contrastive learning to effectively learn representations for information retrieval tasks. The model distinguishes between positive and negative passage pairs, where positive pairs are generated through independent window cropping from the original context (a document), and random token deletion. These approaches ensure that positive pairs share semantic content while exhibiting variation in phrasing, but also occasionally retaining lexical overlap. Negative pairs, on the other hand, are mined using MoCo\cite{he2020momentumcontrastunsupervisedvisual}, a method that builds a queue and uses a slowly changing encoder to generate negative samples. The model updates its document encoder by incorporating an online average of past parameters, ensuring that representations remain consistent across nearby training steps, hence making old representations stored in the queue compatible. This self-supervised training approach enables Contriever to produce dense vector representations for queries and documents, facilitating efficient retrieval and robust generalization across various retrieval tasks without the need for labeled datasets. The model used in this study was further fine-tuned on the English MS MARCO \cite{bajaj2018msmarcohumangenerated}.

\item[Simcse-dist-mpnet-paracrawl-cs-en (SimCSE)\cite{gao2022simcsesimplecontrastivelearning,bednář2023likesmallczechsemantic}.]
SimCSE employs contrastive learning to generate sentence embeddings, using simple dropout-based noise to create positive pairs from the same sentence while drawing negative pairs from other sentences within the batch. This approach trains the model to capture semantic similarities and differences between sentences without relying on large supervised datasets. Positive pairs are formed through augmentation techniques, such as random token deletion, replacement and masking, which introduce variability while preserving the original meaning. We chose to specifically test a model trained on the ParaCrawl \cite{banon-etal-2020-paracrawl} dataset (\verb|SimCSE-Dist-MPNet-ParaCrawl|) as it achieved the highest DaReCzech performance at P@10 in the original work. The model used in this study was pre-trained using an undisclosed Czech dataset from Seznam.cz and distilled on \verb|czeng20-csmono| \cite{kocmi2020announcing} and \verb|Paracrawler v9| \cite{banon-etal-2020-paracrawl}.

\item[BGE Multilingual Gemma2 (Gemma2) \cite{chen2024bgem3embeddingmultilingualmultifunctionality}.]
BGE-Multilingual-Gemma2 is a large-language model based multilingual embedding model. It is directly finetuned using contrastive objective on an undisclosed diverse set of languages and tasks based on \verb|google/gemma-2-9b| model \cite{gemmateam2024gemma2improvingopen}. During evaluation, the prompt used was: "Given a web search query, retrieve relevant passages that answer the query," as outlined in the instructions \footnote{\url{https://huggingface.co/BAAI/bge-multilingual-gemma2}}.

\end{description}

\section{Experimental Setup}

\subsection{DaReCzech Dataset}
\label{sec:dareczech_dataset}
DaReCzech is a Czech dataset designed for text relevance ranking, comprising over 1.6 million query-document pairs. It is divided into Train-big (1.4M pairs for model training), Train-small (97K pairs for model training), Dev (41K pairs), and Test (64K pairs), with no overlap between splits. Each record includes a query, URL, document title, document body text extract (BTE), and a relevance label. Queries are real user inputs, with minor corrections, and the documents are preprocessed to exclude irrelevant sections, ensuring a cleaner representation of content for ranking tasks.

We utilized DaReCzech by selecting test queries along with their associated relevant documents and additional documents to create a 100,000-document sample for indexing. This approach allowed us to maintain a representative document pool without indexing the entire dataset, primarily due to computational and economical overhead. Specifically, for the OpenAI Ada model, embedding generation incurs a cost. This balanced approach enabled a comprehensive evaluation while managing resource expenditure effectively. 
For relevance scores, we classified documents with scores above 0 as relevant and those with a score of 0 as non-relevant, aligning with binary metrics like precision and recall. More about the evaluation criteria can be found in Appendix~\ref{appendix:evaluation}.

\subsection{BM25 grid search}
For fair comparison, we ran a grid search on the development set within our corpus to find the most optimal setting of the BM25's hyperparameters. The performance of the BM25 model was most optimal when the document length normalization parameter $B$ was set to its maximum value of 1.0. This adjustment highlighted the importance of document length normalization in our particular case. The $K_1$ parameter, saturation of term frequency, showed minimal impact on our corpus, suggesting that its tuning had little to no effect on the performance. Based on this experiment, the BM25 hyperparameters were set to $[K_1, B] = [2, 1]$.

\subsection{Dataset Translation}
Some of the models tested were primarily or exclusively trained on English data. To achieve optimal performance and ensure a fair comparison, we applied document-level translation to the DaReCzech corpus, translating it into English using OPUS-MT, a multilingual translation model based on the OPUS corpora \footnote{OPUS-MT translation model: \url{https://huggingface.co/Helsinki-NLP/opus-mt-cs-en}.}. This approach allowed us to evaluate all models in their supported language setting.

\subsection{Segmentation}
For the purpose of our evaluation, we employed a common indexing methodology across all models. For an initial experiment, we indexed documents in two ways: using only a truncated section up to each model’s maximum input limit, and as multiple overlapping segments for longer documents, thus running the evaluation with two separate indices for each model\footnote{As a result, the evaluation with overlapping made the effective number of document represenations in the (single-vector) indices exceed the base count of 100,000 documents. During the retrieval process, all the duplicate versions of documents were removed leaving only the highest rank of the same document.}.  However, since the overlapping approach did not yield any significant improvement, as can be seen in Figures \ref{fig:prec_recall_comparison} and \ref{fig:mrr_ndcg_comparison}, the later tested models were evaluated using non-overlapping segments only. The overlapping segments revealed an inherent bias of DaReCzech, as all the important data were usually concentrated at the beginning of the documents. This was also indicated by our extra analysis in Appendix \ref{appendix:interpretability}, making the cutoff method with no overlap sufficient. The cutoff lengths for each model were derived from the respective model papers, and a stride (if used) was selected to be roughly one-third of the maximum token length for each model as can be seen in Table \ref{tab:experimental_setup}.

\begin{table}[ht]
    \centering
    \
\caption{Overview of Experimental Model Configurations. All models were tested using truncated inputs, with select models additionally tested using segmented document inputs. The Segmenting column specifies the maximum token length and overlap window used for these experiments. OpenAI Ada is an only closed-access model we tested. (*The SPLADE output dimension represents the average number of tokens present in the output vector.).}
     \begin{tabular}{lccccc}
        \hline        
        \textbf{Model} & \textbf{Output Dim.} & \textbf{Max Tokens} & \textbf{Lang.} & \textbf{Segmenting} \\
        \hline
        Splade           & *45.7                & 256                & en    & 256/86    \\
        PLAID            & 128                 & 300                & en    & 300/100   \\
        PLAID-X          & 128                 & 180                & cs    & 180/60    \\
        Contriever       & 768                 & 256                & en    & --        \\
        SimCSE           & 256                 & 128                & cs    & --        \\
        Gemma2         & 3584                & 512                & cs/en & --        \\
        \hline
        OpenAI ada       & 1536                & 8192               & cs    & --        \\
        \hline
    \end{tabular}
    \label{tab:experimental_setup}
\end{table}

\section{Results and Analysis}

\begin{figure}[ht]
    \centering
    \begin{subfigure}[b]{0.49\textwidth}
        \centering
        \includegraphics[width=\textwidth]{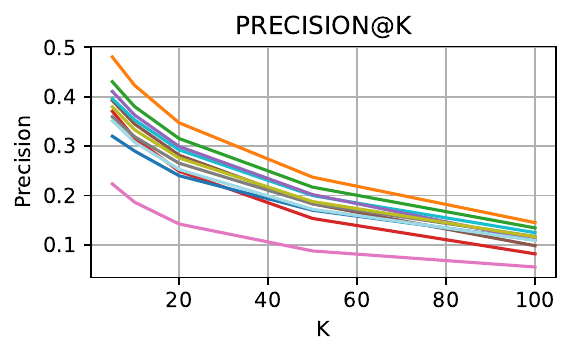}
        \caption{Precision@K}
        \label{fig:prec_k_comparison}
    \end{subfigure}
    \hfill
    \begin{subfigure}[b]{0.49\textwidth}
        \centering
        \includegraphics[width=\textwidth]{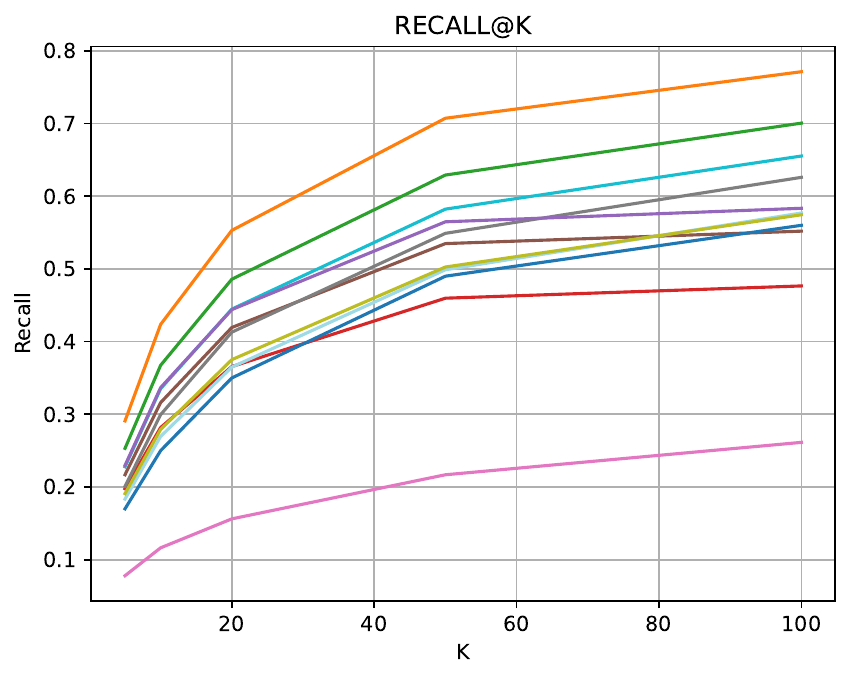}
        \caption{Recall@K}
        \label{fig:recall_k_comparison}
    \end{subfigure}
  
    \begin{subfigure}[b]{0.4\textwidth}
        \centering
        \includegraphics[width=\textwidth]{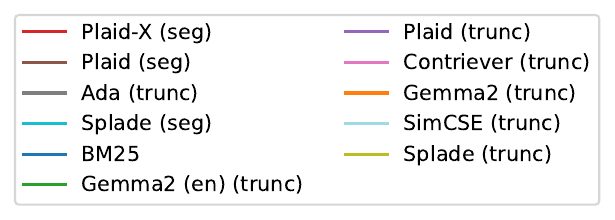}
    \end{subfigure}
    \caption{Comparison of Precision and Recall at different values of $k$.}
    \label{fig:prec_recall_comparison}
\end{figure}

The precision and recall values (Figures \ref{fig:prec_k_comparison},\ref{fig:recall_k_comparison}) for all models across different $k$ values exhibit distinct patterns, with Contriever performing the worst, even below BM25. The stricly top-performing model is Gemma2. Notably, both the Czech and English versions of Gemma2 rank highly, with the Czech model showing a slight advantage in performance. Beneath Gemma2, the best results come from the PLAID models. However, segmenting the documents with these models demonstrates a decline in both precision and recall as $k$ increases, possibly due to an accumulation of irrelevant information from segment-level retrievals impacting the overall ranking quality\footnote{The evaluation was conducted in two ways for some models. In one the retrieval function retrieve(k) was called for each tested value of $k$ separately and in the other the highest tested value of $k$ was chosen and then the results cut off down to the tested $k$ value. This was done to especially examine the PLAID retrieval implementation, which determines different hyperparameters for its approximate nearest-neighbor search for different $k$ value. This however did not show any significant change in the tested metrics (in these cases the better results were kept).}.

\begin{figure}[H]
    \centering
    \begin{subfigure}[b]{0.48\textwidth}
        \centering
        \includegraphics[width=\textwidth]{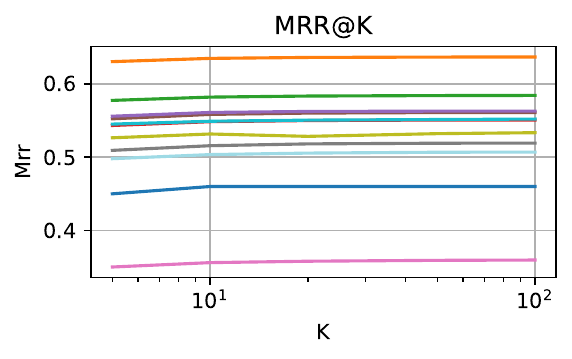}
        \caption{MRR@K}
        \label{fig:mrr_comparison}   \end{subfigure}
    \hfill
    \begin{subfigure}[b]{0.48\textwidth}
        \centering
        \includegraphics[width=\textwidth]{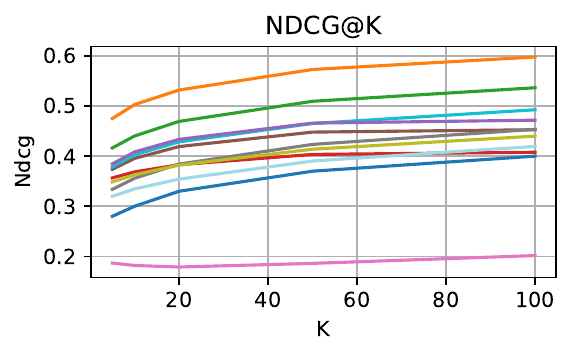}
        \caption{NDCG@K}
        \label{fig:ndcg_comparison}
    \end{subfigure}
    \begin{subfigure}[b]{0.4\textwidth}
        \centering
        \includegraphics[width=\textwidth]{plots/comparisons/legend_only.pdf}
    \end{subfigure}
    \caption{Comparison of MRR, NDCG at different values of $k$.}
    \label{fig:mrr_ndcg_comparison}
\end{figure}

These observed trends are even more evident in the full MRR and NDCG metrics (Figures \ref{fig:mrr_comparison} \ref{fig:ndcg_comparison}), where the differences among models are more pronounced. In the MRR graphs, nearly consistent performance across different $k$ values indicates that while the general retrieval ability remains stable, the ranking quality of results varies significantly between models. The superior performance of Gemma2 and the relative weaknesses of Contriever are reflected here, reinforcing the patterns observed in the precision and recall figures. This alignment suggests that models with higher precision and recall also exhibit better ordering and ranking capabilities, as demonstrated by their MRR and NDCG scores.

\begin{figure}[ht]
    \centering
    \begin{subfigure}[b]{0.45\textwidth} 
        \centering
        \includegraphics[width=\textwidth]{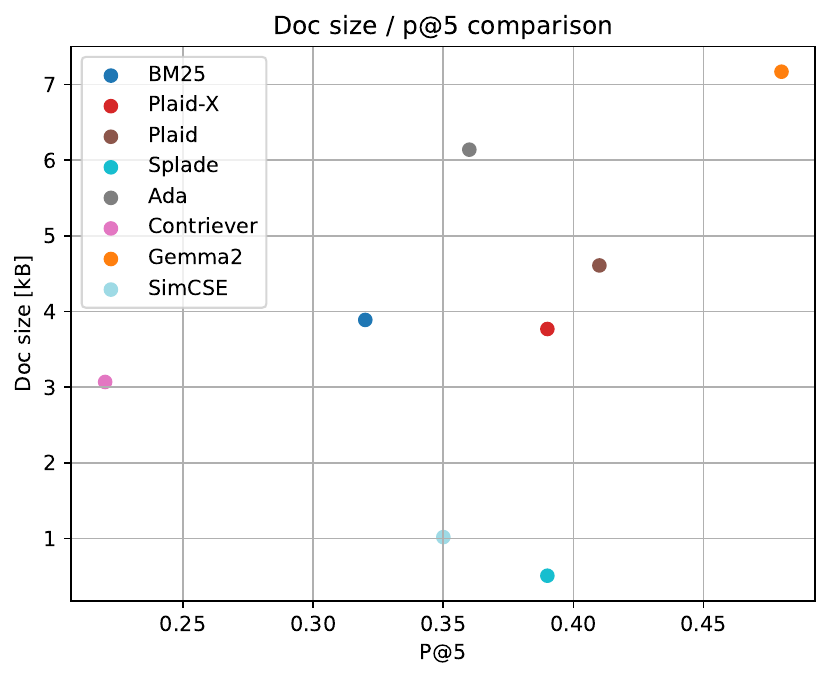}
        \caption{Doc repr. size / P@5 comparison}
        \label{fig:docsize_v_precision}
    \end{subfigure}
    \hfill
    \begin{subfigure}[b]{0.52\textwidth}
        \centering
        \includegraphics[width=\textwidth]{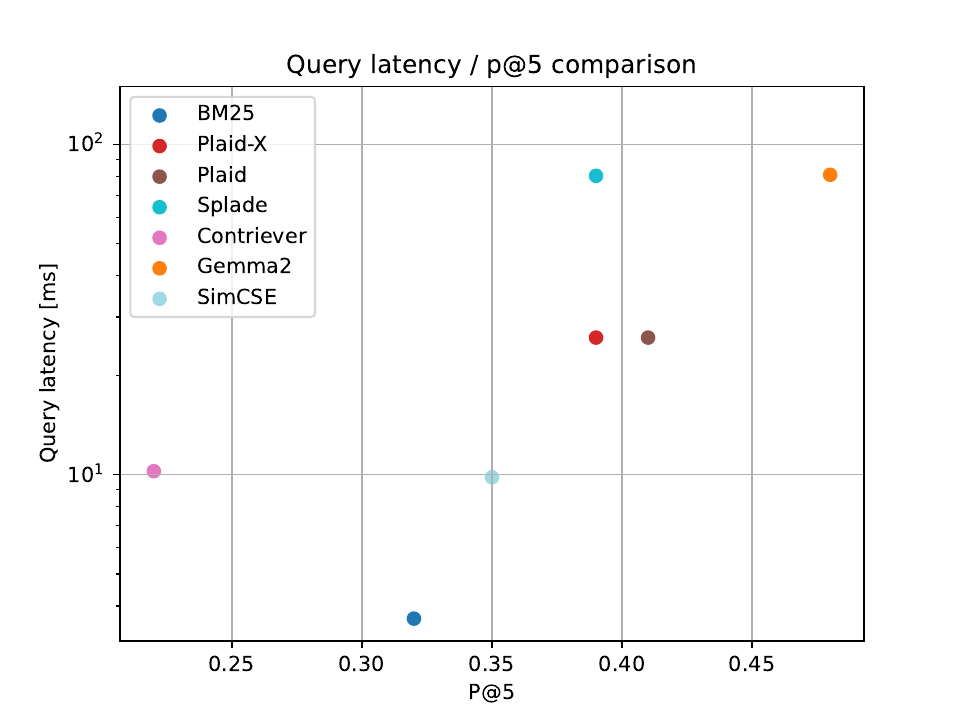}
        \caption{Query latency / P@5 comparison}
        \label{fig:latency_v_precision}
    \end{subfigure}
    \caption{Doc size, query latency in relation to P@5.}
    \label{fig:comparison}
\end{figure}

Figure~\ref{fig:docsize_v_precision} demonstrates the trade-off between document size (estimated by averaging the size of each index by the number of indexed documents) and retrieval precision. As anticipated, BM25 maintains a compact document representation but exhibits a low Precision@5 performance, with Contriever faring even worse. PLAID-X achieves modest gains over BM25, with a smaller index size per document due to a restrictive 180-token limit. SPLADE, while comparable to PLAID-X in precision, maintains a much smaller index thanks to its sparse nature \footnote{Some models achieve similar or even significantly smaller index sizes compared to BM25 due to truncation; BM25 indexed entire documents without truncation, while many other models were limited to a few hundred tokens per document. This limit, especially for longer documents, led to reduced overall index sizes.}. The original PLAID model, without cross-lingual settings, slightly outperforms both PLAID-X and SPLADE, though it incurs a larger index size due to a higher token limit of 300. OpenAI's Ada model struggles to compete, hindered by its large embedding dimension, resulting in a substantial index size that does not justify its middling performance. The Gemma2 model emerges as the top performer, albeit with the largest embedding size, indicating a trade-off between high retrieval accuracy and storage requirements. Such result is aligned with observations in \cite{neelakantan2022text}, where authors demonstrate that embedding performance tends to scale with model size and embedding dimension.

An analysis of query latency in Figure~\ref{fig:latency_v_precision} shows that BM25 achieves the fastest query times, which aligns with its straightforward term-matching approach. Contriever and SimCSE, both using single-vector embeddings and cosine similarity, follow closely. The PLAID-X and PLAID models exhibit slightly longer latencies, likely due to their multi-stage retrieval process, which involves candidate selection and more complex ranking steps, contributing to a moderate increase in query time. SPLADE and Gemma2 are slower still; SPLADE’s sparse representation requires additional computation to dynamically calculate sparse scores, while Gemma2’s high-dimensional embeddings impose added processing overhead. These patterns suggest that models with multi-stage or complex scoring mechanisms naturally incur higher latency compared to more direct embedding or term-based approaches.

\begin{figure}[H]
    \centering
    \begin{subfigure}[b]{0.48\textwidth}
        \centering
        \includegraphics[width=\textwidth]{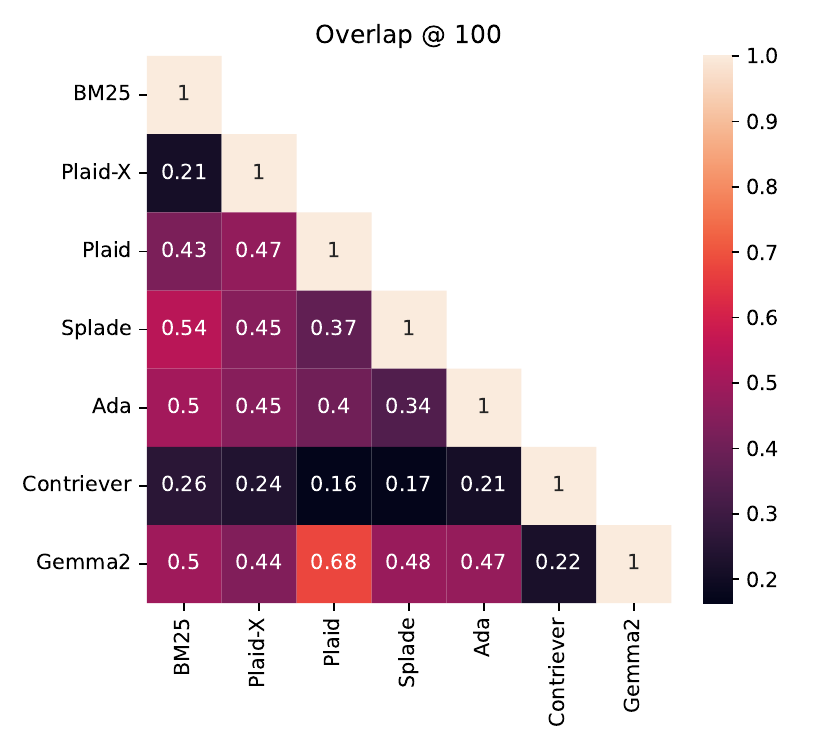}
        \caption{Overlap @ 100}
    \end{subfigure}
    \hfill
    \begin{subfigure}[b]{0.48\textwidth}
        \centering
        \includegraphics[width=\textwidth]{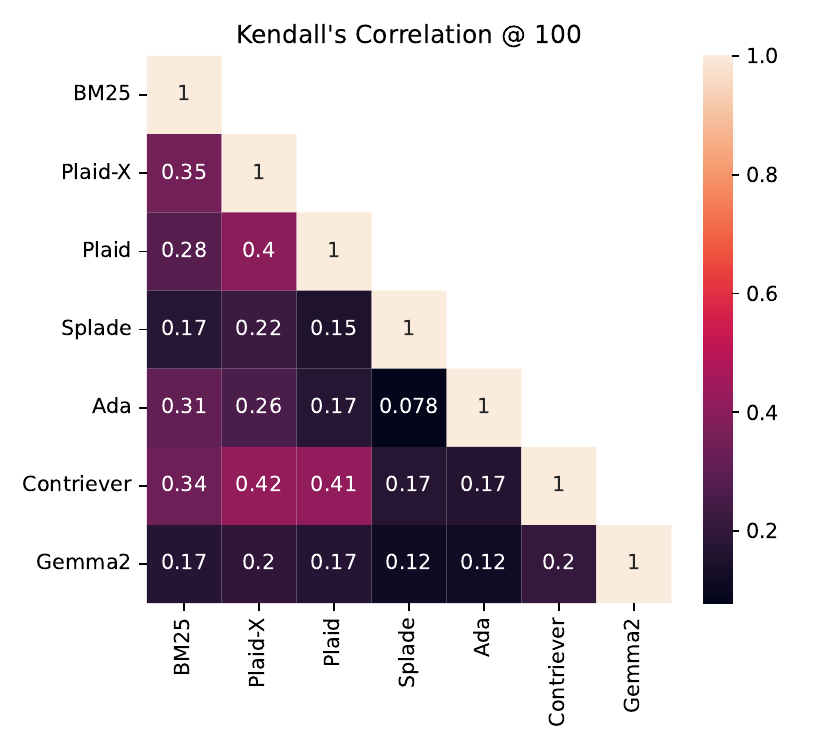}
        \caption{Kendall's Correlation @ 100}
    \end{subfigure}
    \caption{Pairwise overlap and correlation of overlapped items in top-100 responses of different IR systems.}
    \label{fig:overlaps_correlation}
    % data source: https://drive.google.com/drive/folders/1NFu29kHL0Ygs_E7dPX573TYH66iur7fh?usp=sharing
    % visualizations source notebook:https://colab.research.google.com/drive/1YCC8E2MSPWnoOi1pYQirk2moXBwA0yI3?usp=sharing
\end{figure}

Regarding the overlap among the models, as depicted in Figure~\ref{fig:overlaps_correlation}, Contriever consistently exhibits the lowest overlap scores across comparisons with other models, a finding that aligns well with its previously observed underperformance in Figures \ref{fig:prec_k_comparison} and \ref{fig:recall_k_comparison}. Notably, PLAID and PLAID-X display a high degree of overlap and strong Kendall $\tau$ correlation, likely attributable to their shared architecture and training approach, with PLAID-X being a multilingual adaptation of PLAID. Interestingly, we also observe a notably high overlap value between PLAID and GEMMA, which could be attributed to GEMMA's training on diverse multilingual data that likely includes features common to PLAID's retrieval methodology. 

\section{Conclusion}
In this paper, we evaluated 7 off-the-shelf information retrieval models on the DaReCzech corpus, comparing their performance against the traditional BM25 approach. The goal was to identify the most effective model for information retrieval in Czech.

Our findings showed that Gemma2 consistently delivered the best precision and recall metrics across various $k$ values, with the Czech version slightly outperforming the English one. However, its high retrieval accuracy came with a large index size due to high-dimensional embeddings exceeding even the multi-vector models. In contrast, BM25 and Contriever exhibited the poorest performance, with Contriever notably underperforming and struggling to match BM25’s baseline.

SPLADE and the PLAID models offered a balance between performance and efficiency. SPLADE's sparse representation resulted in the smallest index size, making it suitable for resource-constrained applications, while the PLAID models, especially the original, provided higher precision with modest increases in index size. The ColBERT-based models performed well unsegmented, but segmenting for long documents led to a decrease in performance as $k$ increased.

For Czech-language IR tasks, Gemma2 is recommended if accuracy is the top priority and storage is manageable. SPLADE is a practical choice when memory efficiency is crucial, and PLAID/PLAID-X offer a middle ground, particularly with token limit adjustments. This study underscores the trade-offs between model complexity, storage, and retrieval quality, guiding suitable model selection for Czech-language IR.

\begin{credits}
\subsubsection{\ackname} This work was supported by project Ministry of Culture of the Czech Republic through NAKI III project semANT, grant. no DH23P03OVV060, Horizon EU programme through project ELOQUENCE, grant no. 101135916, and by the Ministry of Education, Youth and Sports of the Czech Republic through the e-INFRA CZ (ID:90254).

\subsubsection{\discintname}
The authors have no competing interests to declare that are relevant to the content of this article.
\end{credits}

\bibliographystyle{splncs04}
\bibliography{references}

\begin{appendix}
\section{Evaluation Process}
\subsection{Evaluation Metrics}
\label{appendix:evaluation}
To assess the performance of these IR models, we employ a range of standard evaluation metrics:

\subsubsection{Precision}
Precision quantifies the accuracy of relevant documents in the retrieved set and is defined as:
\begin{equation}
\text{Precision} = \frac{|\{\text{relevant documents}\} \cap \{\text{retrieved documents}\}|}{|\{\text{retrieved documents}\}|}
\end{equation}

\subsubsection{Recall}
Recall measures the ability of the model to retrieve all relevant documents from the corpus and is given by:
\begin{equation}
\text{Recall} = \frac{|\{\text{relevant documents}\} \cap \{\text{retrieved documents}\}|}{|\{\text{relevant documents in corpus}\}|}
\end{equation}

\subsubsection{MRR (Mean Reciprocal Rank)}
Mean Reciprocal Rank (MRR) evaluates the ranking quality by taking the mean of the reciprocal ranks of the first relevant document for each query:
\begin{equation}
\text{MRR} = \frac{1}{|Q|} \sum_{i=1}^{|Q|} \frac{1}{\text{rank}_i}
\end{equation}
where $\text{rank}_i$ is the position of the first relevant document for the $i$-th query and $|Q|$ is the total number of queries.

\subsubsection{MAP (Mean Average Precision)}
Mean Average Precision (MAP) calculates the average precision for each query and averages these scores across all queries, thereby reflecting the model’s ranking consistency. For a given query $q$, the average precision is:
\begin{equation}
\text{AP}_q = \frac{1}{|\{\text{relevant documents for } q\}|} \sum_{k=1}^{N} \text{Precision}(k) \cdot \text{rel}(k)
\end{equation}
where $N$ is the total number of documents, $\text{Precision}(k)$ is the precision at rank $k$, and $\text{rel}(k)$ is a binary indicator of relevance at rank $k$. MAP is then:
\begin{equation}
\text{MAP} = \frac{1}{|Q|} \sum_{q=1}^{|Q|} \text{AP}_q.
\end{equation}

\subsubsection{nDCG (Normalized Discounted Cumulative Gain)}
Normalized Discounted Cumulative Gain (nDCG) evaluates the ranked list’s quality by considering the position of relevant documents in the ranking. For a query $q$, nDCG at rank $p$ is calculated as:
\begin{equation}
\text{DCG}_p = \sum_{k=1}^{p} \frac{2^{\text{rel}(k)} - 1}{\log_2(k+1)}
\end{equation}
\begin{equation}
\text{nDCG}_p = \frac{\text{DCG}_p}{\text{IDCG}_p}
\end{equation}
where $\text{rel}(k)$ is the relevance score of the document at rank $k$, and $\text{IDCG}_p$ is the ideal DCG, obtained by sorting documents in the perfect order of relevance.

\subsection{Additional Evaluation Criteria}
In addition to the standard metrics, we also consider two specific criteria:

\subsubsection{Representation Size}
Examining the memory footprint required by each model's document representations (measured per document as kB/doc),

\subsubsection{Query Latency}
Query latency refers to the duration taken by an information retrieval system to retrieve and present relevant documents in response to a given query.

\subsubsection{Kendall's $\tau$ Rank Correlation and Lexical Overlap}
assessing the consistency of ranking across the models on the top 100 retrieved results for each query - helps understanding how well the models agree on the most relevant documents

\section{BM25 Hyperparameter Tuning}
\label{appendix:baseline}
\begin{figure}[ht]
    \centering 
    \includegraphics[width=0.6\textwidth]{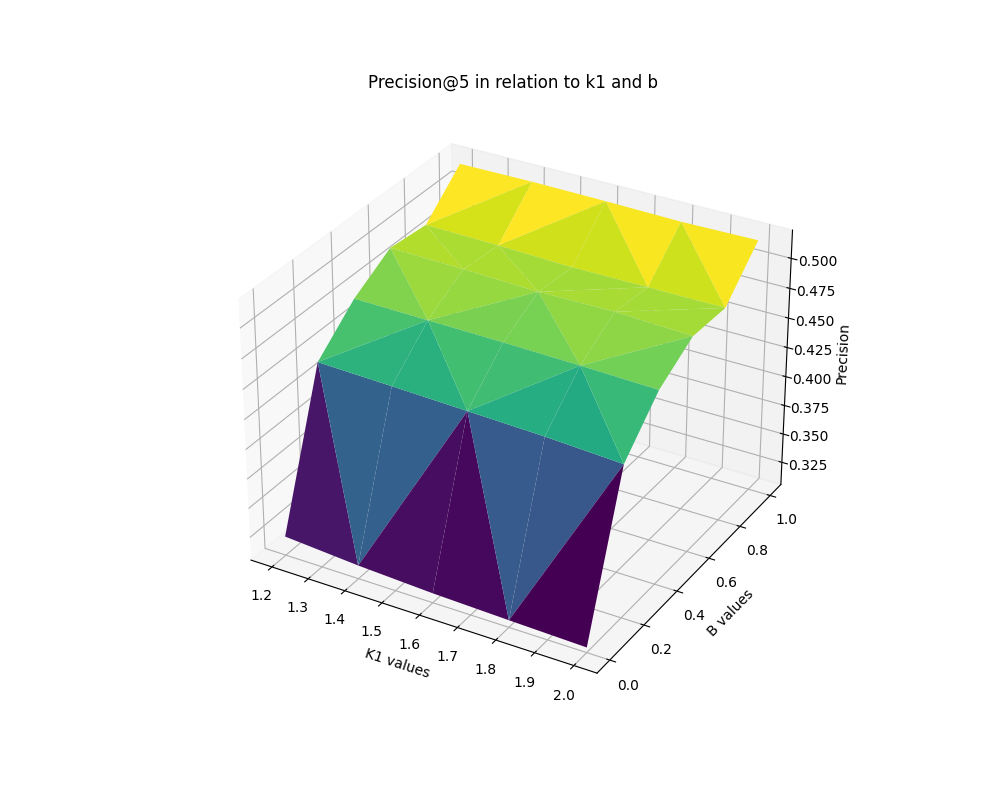}
    \caption{BM25 hyperparameters grid search.}
    \label{fig:grid_search}
\end{figure}

We perform a BM25 grid search to tune the K1 and B parameters for optimal results on the corpus. The results from the grid search are visualized in Figure~\ref{fig:grid_search}.

\section{ColBERT's  Token-level Focus}
\label{appendix:interpretability}
To estimate which parts of the document are important, the study analyzed ColBERTv2, a model that uses a multi-vector approach, where each token in a document is represented by a separate vector. By examining the vectors of the retrieved documents, tokens with the most interaction with the query were identified. This might indicate which specific parts of the documents were most relevant to the query and contributed to the retrieval process for the given document.

This experiment examined two modes for the document-query scores (samples with scores aggregated from Colbert's similarity matrix through max-pooling over query token representations (in contrast with the original Colbert's MaxSim operation which computes max-pooling over document token representations) can be seen below in Figure~\ref{fig:colbert_interpretability}) visualized as a probability distribution using the softmax function with the brighter color denoting higher similarity score):  

MaxSim operation: particularly chosen as it reflects how the model selects positive document-query pairs, and identifies the best score for each query token with the highest scoring document token, highlighting the most significant interactions. 
\begin{figure}[ht]
    \centering 
    \includegraphics[width=0.9\textwidth]{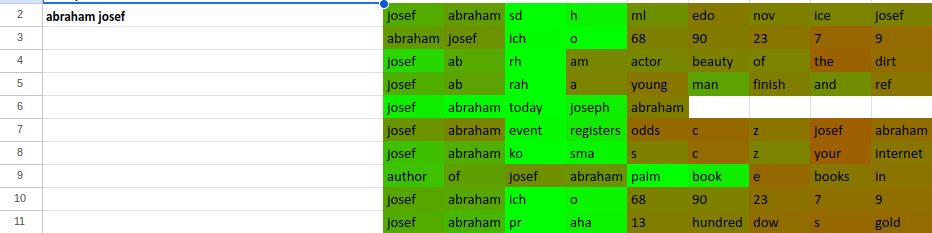}
    \caption{Colbert interpretability.}
    \label{fig:colbert_interpretability}
\end{figure}
\end{appendix}

\end{document}